\documentclass[aps,prb,twocolumn,superscriptaddress,amsmath,amssymb]{revtex4}

\usepackage{amssymb}
\usepackage[dvips]{graphicx}
\usepackage{bm}
\usepackage{epsfig}
\usepackage[ansinew]{inputenc}

\begin{document}

\title{Pressure effects on the structural and superconducting transitions in La$_3$Co$_4$Sn$_{13}$}

\author{L. Mendon\c{c}a-Ferreira}
\affiliation{Centro de Ci\^{e}ncias Naturais e Humanas, Universidade Federal do ABC, 09210-170, Santo Andr\'{e}, SP, Brazil}

\author{F. B. Carneiro}
\affiliation{Centro Brasileiro de Pesquisas F\'{\i}sicas, 22290-180, Rio de Janeiro, RJ, Brazil}

\author{M. B. Fontes}
\affiliation{Centro Brasileiro de Pesquisas F\'{\i}sicas, 22290-180, Rio de Janeiro, RJ, Brazil}

\author{E. Baggio-Saitovitch}
\affiliation{Centro Brasileiro de Pesquisas F\'{\i}sicas, 22290-180, Rio de Janeiro, RJ, Brazil}

%\author{M. A. Continentino}
%\affiliation{Centro Brasileiro de Pesquisas F\'{\i}sicas, 22290-180, Rio de Janeiro, RJ, Brazil}

\author{L. S. I. Veiga}
\altaffiliation{Present address: London Centre for Nanotechnology and Department of Physics and Astronomy, University College London, Gower Street, London, WC1E 6BT, United Kingdom}
\affiliation{Deutsches Elektronen Synchrotron (DESY), Notkestrasse 85, 22603, Hamburg, Germany}

\author{J. R. L. Mardegan}
\affiliation{Deutsches Elektronen Synchrotron (DESY), Notkestrasse 85, 22603, Hamburg, Germany}
\affiliation{Swiss Light Source, Paul Scherrer Institut, 5232 Villigen PSI, Switzerland}

\author{J. Strempfer}
\affiliation{Deutsches Elektronen Synchrotron (DESY), Notkestrasse 85, 22603, Hamburg, Germany}

\author{M. M. Piva}
\affiliation{Instituto de F\'{\i}sica ``Gleb Wataghin", UNICAMP, 13083-859, Campinas, SP, Brazil}

\author{P. G. Pagliuso}
\affiliation{Instituto de F\'{\i}sica ``Gleb Wataghin", UNICAMP, 13083-859, Campinas, SP, Brazil}

\author{R. D. dos Reis}
\affiliation{Laborat\'{o}rio Nacional de Luz S\'{\i}ncrotron (LNLS), Centro Nacional de Pesquisa em Energia e Materiais (CNPEM), CEP 13083-970, Campinas, SP, Brazil}

\author{E. M. Bittar}
\email{bittar@cbpf.br}
\affiliation{Centro Brasileiro de Pesquisas F\'{\i}sicas, 22290-180, Rio de Janeiro, RJ, Brazil}

\begin{abstract}
La$_3$Co$_4$Sn$_{13}$ is a superconducting material with transition temperature at $T_{c}=2.70$ K, which presents a superlattice structural transition at $T^*\simeq150$ K, a common feature for this class of compounds. However, for this material, it is not clear that at $T^*$ the lattice distortions arise from a charge density wave (CDW) or from a distinct microscopic origin. Interestingly, it has been suggested in isostructural non-magnetic intermetallic compounds that $T^*$ can be suppressed to zero temperature, by combining chemical and external pressure, and a quantum critical point is argued to be observed near these critical doping/pressure. Our study shows that application of pressure on single-crystalline La$_3$Co$_4$Sn$_{13}$ enhances $T_c$ and decreases $T^*$. We observe thermal hysteresis loops for cooling/heating cycles around $T^*$ for $P\gtrsim0.6$ GPa, in electrical resistivity measurements, which are not seen in x-ray diffraction data. The hysteresis in electrical measurements may be due to the pinning of the CDW phase to impurities/defects, while the superlattice structural transition maintains its ambient pressure second-order transition nature under pressure. From our experiments we estimate that $T^*$ vanishes at around 5.5 GPa, though no quantum critical behavior is observed up to 2.53 GPa.
\end{abstract}

\maketitle

\section{Introduction}
There has been a revived interest in the R$_3$M$_4$X$_{13}$ (3-4-13) (R = rare-earth or alkaline-earth element; M = transition metal and X = groups-13,14 element) compounds due to their diverse physical properties, which include antiferromagnetic \cite{amelia,leticie,giles,sato}, superconducting \cite{sato,morosange}, strong electronic correlations \cite{sato,jacke} and semiconducting \cite{ge,morosange} behavior. The crystal structure of this 3-4-13 series, at room temperature, is the cubic Yb$_{3}$Rh$_{4}$Sn$_{13}$ type structure ($Pm\bar{3}n$ space group) \cite{hodeau}. The cubo-octahedral R site has a small distortion, resulting in a local tetragonal symmetry \cite{oscar}. For M = Co and X = Sn, all known compounds undergo a superlattice structural transition at a temperature $T^*$, which doubles the lattice parameter in respect to the higher temperature phase \cite{oscar}, with propagation vector of $\textbf{q}={\{(0.5 \ 0.5 \ 0),(0.5 \ 0 \ 0.5),(0 \ 0.5 \ 0.5)\}}$ \cite{neutrons}. This superlattice structural transition was also observed for the Sr$_3$Ir$_4$Sn$_{13}$ superconductor and the existence of a charge density wave (CDW) state below $T^*\simeq147$ K was reported \cite{qcp3413}. By combining Ca$^{2+}$ substitution and application of external pressure $P$, $T^*$ is suppressed to zero, suggesting a superlattice quantum phase transition at zero temperature \cite{qcp3413}, analogous to those near a magnetic \cite{mucio,RevLohneysen}, superconducting \cite{scqcp} and ferroelectric \cite{ferroqcp} instabilities. Later, a structural quantum critical point was also described in the Sr$_{3-x}$Ca$_x$Rh$_4$Sn$_{13}$ system at ambient pressure \cite{qcp34132}.

Due to its similarity with the Sr$_{3-x}$Ca$_x$(Rh,Ir)$_4$Sn$_{13}$ series of materials, the La$_3$Co$_4$Sn$_{13}$ compound, presenting a superconducting transition temperature at $T_{c}=2.70$ K, has recently attracted attention \cite{cecosn,lacosnX,liu}. The superconducting state is argued to be $s$-wave, in the strong-coupling regime \cite{DFT,BCS}, such as others 3-4-13 superconductors \cite{StrongC1,StrongC2,StrongC3}. The superlattice structural transition of La$_3$Co$_4$Sn$_{13}$ occurs at $T^*\simeq150$ K and whether this transition is of first or second-order is still under debate \cite{liu,lacosnX}. It is also controversial whether at $T^*$ the lattice distortions arise from a conventional CDW \cite{liu,lacosnX2}.

Pressure effects studies on superconductors are of great interest since it gives insights on the microscopic superconducting mechanism and may also induce interesting physical phenomena that might arise out of ambient conditions. For simple metal superconductors it is generally expected that $T_c$ decreases with pressure. This is because the density of states at the Fermi energy [$N(E_f)$] and the effective attractive pairing interaction ($V_{eff}$), which are both related to $T_c$ through $T_c\propto e^{-1/[N(E_f)V_{eff}]}$, vary with pressure. While $N(E_f)$ usually decreases with the reduction of the unit cell volume, $V_{eff}$ is significantly affected due to the difficulty for the crystal lattice to couple with the electrons \cite{Schilling,Chu}. Oddly, it was observed that for a polycrystalline La$_3$Co$_4$Sn$_{13}$ sample ($T_{c}=1.95$ K at ambient pressure) $T_{c}$ increases linearly with applied pressure at a rate of $dT_{c}/dP\sim0.03$ K/GPa \cite{LaP}, in contrast to M = Rh \cite{LaP} and Ru \cite{LaP2} in which $T_{c}$ decreases with pressure. An increase of $T_{c}$ up to 5.1 K, for 10\% In substitution, was also reported, though application of pressure in La$_3$Co$_4$Sn$_{11.7}$In$_{1.3}$ decreased $T_{c}$ by -0.3 K/GPa \cite{neha}. While the enhancement of $T_{c}$ against pressure is attributed to the decrease of the structural instability \cite{LaP}, the results for In substitution, based on theoretical calculations, is argued to be related to modifications in band structure and density of states \cite{neha}.

In this work we present our investigation of pressure effects on single-crystalline superconducting La$_3$Co$_4$Sn$_{13}$ compound, via electrical resistivity and x-ray diffraction. We observe a positive variance of $T_{c}$ under pressure, with a faster rate than for polycrystalline samples. Additionally, a thermal hysteresis loop around the superlattice structural transition is seen in electrical resistivity experiments for $P\gtrsim0.6$ GPa, presumably due to pinning of a partially gapped CDW phase, which sets in at $T^*$. This feature was not previously reported, giving evidence for a concomitant CDW at $T^*$, and may be a common characteristic of some 3-4-13 materials. We also show the decrease of this second-order superlattice structural transition with pressure and estimate that it vanishes at $P\sim5.5$ GPa, however, with no indication of quantum critical behavior up to $P=2.53$ GPa.

\section{Experimental details}

Flux-grown single crystals of La$_3$Co$_4$Sn$_{13}$ were synthesized using Sn excess \cite{cecosn}. Phase purity was checked by x-ray diffraction (XRD) on powdered crystals (not shown). Ambient pressure synchrotron XRD data ($E=8.33$ keV) at cooling and heating cycles were collected for a single crystalline sample ($\sim2\times1\times1$ mm$^3$), cut and polished to achieve a flat and shiny surface perpendicular to the $[110]$ direction at beamline P09 at PETRA III (DESY/Germany) \cite{desy}. Temperature dependent DC electrical resistivity measurements were performed in a Quantum Design DynaCool PPMS, by means of the conventional four-contact configuration. For experiments under hydrostatic pressure, several samples were screened for Sn inclusions. We used a clamp-type Cu-Be cell, with silicon oil as pressure transmitting medium. Lead was used as pressure manometer. Synchrotron XRD measurements under pressure were performed at beamline XDS at UVX (LNLS/Brazil) \cite{xds}. The sample/diamond anvil cell (DAC) was placed in the cold-finger of a He cryostat. The powder patterns were collected with an area detector (MAR225) and the two-dimensional (2D) images were integrated to provide intensity as a function of 2$\theta$ using the FIT2D software \cite{xrhp}. Due to the DAC limited angular scattering range (25$^\circ$ of scattering angle 2$\theta$), the beam was tuned to $E=20$ keV in order to detect a significant number of Bragg peaks. Hydrostatic pressure was generated using a DAC with diamond anvils of 900 $\mu$m culet size and silicon oil was used as pressure media. To calibrate \textit{in situ} the pressure, ruby spheres were loaded with the fine powdered samples in the stainless steel gasket specimen chamber.

\section{Results}

\begin{figure}
\begin{center}
\includegraphics[width=0.5\textwidth]{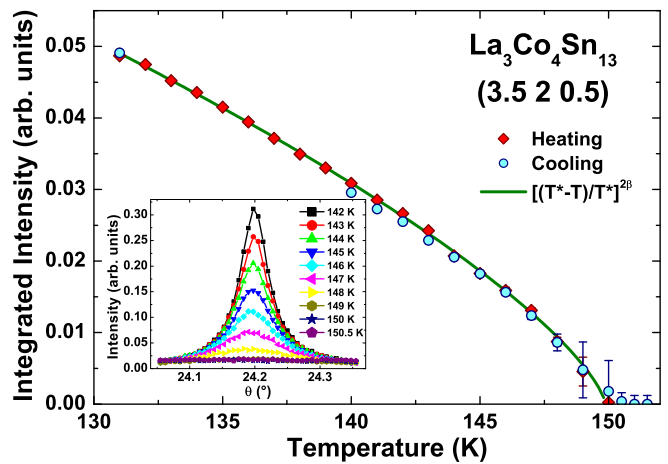}
\end{center}
\caption{Temperature dependence of the (3.5 2 0.5) superlattice reflection at ambient pressure for cooling (circles) and heating (diamonds) cycle. The green solid line is a fitting using a power law to determine $T^*$. The inset shows the rocking curves around the superlattice reflection measured for selected temperatures during the heating cycle.}
\label{FigSXRD}
\end{figure}

Figure \ref{FigSXRD} shows the temperature dependence, upon heating up and cooling down the sample, of the integrated intensity for the superstructure reflection (3.5 2 0.5) (at ambient pressure) of La$_3$Co$_4$Sn$_{13}$. A continuous decrease of the superstructure peak and no appreciable thermal hysteresis, within the instrument resolution, indicate a second-order phase transition at $T^*\simeq150$ K, as in earlier reports \cite{lacosnX,lacosnX2}. In addition, the inset exhibits the rocking curves around the superstructure peak used to calculate the integrated intensities. In order to gain more insight into the physical properties of this structural distortion, the temperature dependent data was fitted by a power-law expression $[(T^*-T)/T^*]^{2\beta}$ yielding $T^*=150.0(1)$ K and a critical exponent $\beta=0.36(1)$. The critical exponent for our single crystal suggests a three-dimensional character of the structural distortion, in good agreement to what is observed on single crystalline Sr$_3$Ir$_4$Sn$_{13}$ \cite{sr3ir4sn13}.

The temperature dependent electrical resistivity $\rho(T)$ of our La$_{3}$Co$_{4}$Sn$_{13}$ single crystals, at ambient pressure, show metallic behavior down to low temperatures, until the superconducting transition takes place at $T_{c}\simeq2.7$ K. Additionally, a small kink due to the superlattice structural transition is observed at $T^*\simeq146$ K, consistent with our synchrotron XRD data. No thermal hysteresis was also observed for $\rho(T)$ and specific heat measurements at $P=0$. These data are consistent with previous works \cite{cecosn,lacosnX} and are not shown here.

\begin{figure}
\begin{center}
\includegraphics[width=0.5\textwidth]{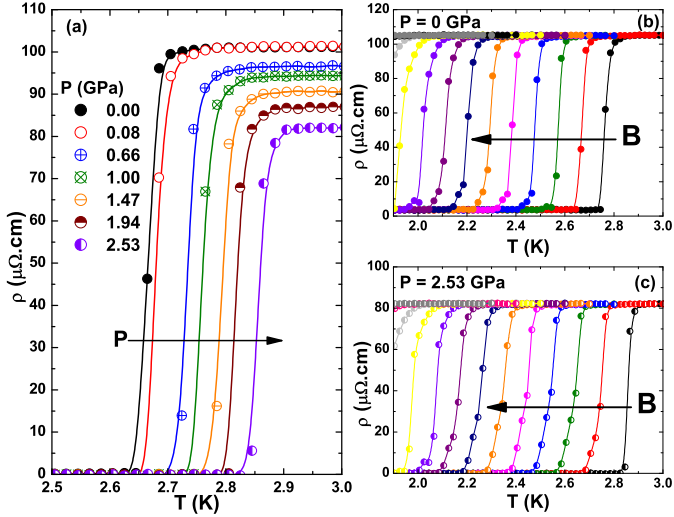}
\end{center}
\caption{(a) Representative pressure-dependent electrical resistivity curves as a function of temperature of the superconducting transition temperature $T_{c}$ in La$_{3}$Co$_4$Sn$_{13}$. The arrows indicate the direction of increasing applied pressure. $B$ dependence of $T_c$ at (b) $P=0$ GPa and (c) $P=2.53$ GPa. $B$ step is 0.05 T (from 0 up to the maximum value of $B=0.6$ T) and the arrows show the direction of the increasing magnetic field. Solid lines are guides to the eyes.}
\label{FigRhoTcP}
\end{figure}

\begin{figure}
\begin{center}
\includegraphics[width=0.5\textwidth]{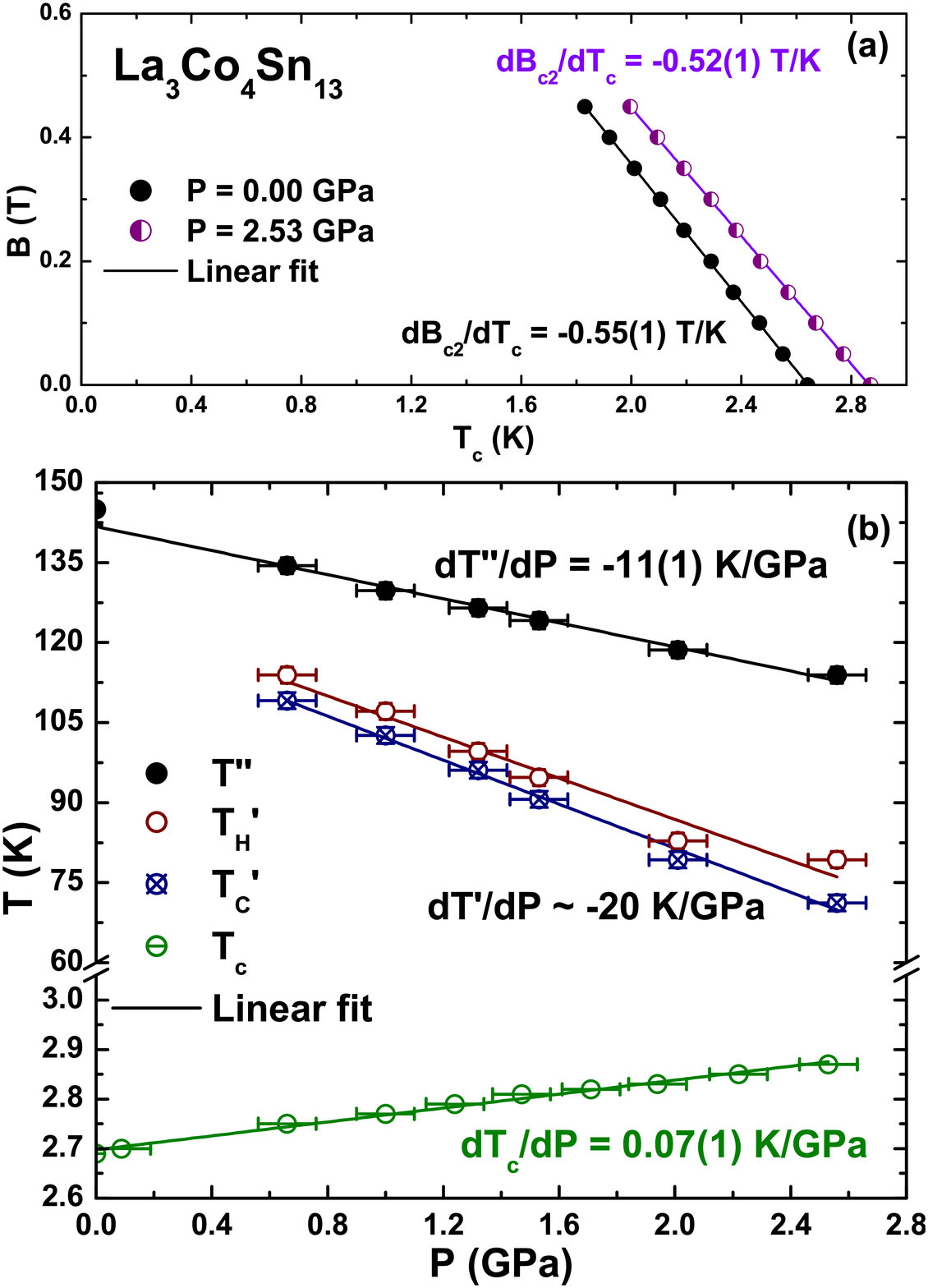}
\end{center}
\caption{(a) Temperature-field phase diagram for La$_{3}$Co$_4$Sn$_{13}$ showing $T_{c}$. (b) Pressure-temperature phase diagram for La$_{3}$Co$_4$Sn$_{13}$ showing the pressure dependence of the superlattice structural transition $T''$, the low temperature hump-like feature onset $T'$ on cooling $T_C'$ and heating $T_H'$, seen in the electrical resistivity data, and the superconducting transition temperature $T_{c}$. Straight lines are linear fits. }
\label{Fig5}
\end{figure}

Representative curves of the pressure evolution of $T_{c}$ in La$_{3}$Co$_{4}$Sn$_{13}$ for our $\rho(T)$ data are presented in Fig. \ref{FigRhoTcP}(a). A sharp transition near $T_{c}$ is observed, which does not broaden with pressure. Similarly to polycrystalline La$_3$Co$_4$Sn$_{13}$, our single crystalline samples also exhibit an enhancement of $T_{c}$ with pressure, increasing linearly at a rate of $dT_{c}/dP\sim0.07(1)$ K/GPa [Fig. \ref{Fig5}(b)], more than double of the polycrystalline material rate \cite{LaP}, but one order of magnitude lower than observed for the initial $dT_{c}/dP$ slope in Ca$_3$Ir$_4$Sn$_{13}$ \cite{arxivca}. As seen in the case for Ca$_3$Ir$_4$Sn$_{13}$ \cite{arxivca}, the residual resistivity $\rho_0$ decreases under pressure. By relaxing the pressure, $\rho_0$ increases again (not shown), attesting that in this pressure range the material is in its elastic regime and only electronic changes are observed.

The magnetic field ($B$) dependence of $T_{c}$ at $P=0$ and 2.53 GPa can be seen in Figs. \ref{FigRhoTcP}(b) and \ref{FigRhoTcP}(c), respectively. A decrease of $T_{c}$ is observed due to the magnetic field pair breaking effect. From these data we construct the temperature-field phase diagram presented in Fig. \ref{Fig5}(a), where $B_{c2}$ is the upper critical field. From the $dB_{c2}/dT$ slope near $T_c$ we estimate $B_{c2}(0)\sim1.0$ T, based on
the Werthamer-Helfand-Hohenberg theory for a BCS superconductor with isotropic gap $[B_{c2}(0)\sim0.7T_c|\frac{dB_{c2}}{dT}|_{T=T_c}]$ \cite{whh}. The coherence length can be calculated from the expression $\xi=(\frac{\Phi_0}{2\pi B_{c2}})^{1/2}$, where the flux quantum is $\Phi_0=2.068\times10^{-15}$ Tm$^2$, giving $\xi=18$ nm. These values are in good agreement with previous reports \cite{LaP,LaP2,neha}. Since the Pauli limiting field is $B_{Pauli}=1.83T_c$, we observe that La$_3$Co$_4$Sn$_{13}$ is an orbital limited superconductor. For pressures up to 2.53 GPa no appreciable change to the product $T_c|\frac{dB_{c2}}{dT}|_{T=T_c}$ is seen, thus the superconducting state properties does not alter significantly in this pressure range.

\begin{figure}
\begin{center}
\includegraphics[width=0.5\textwidth]{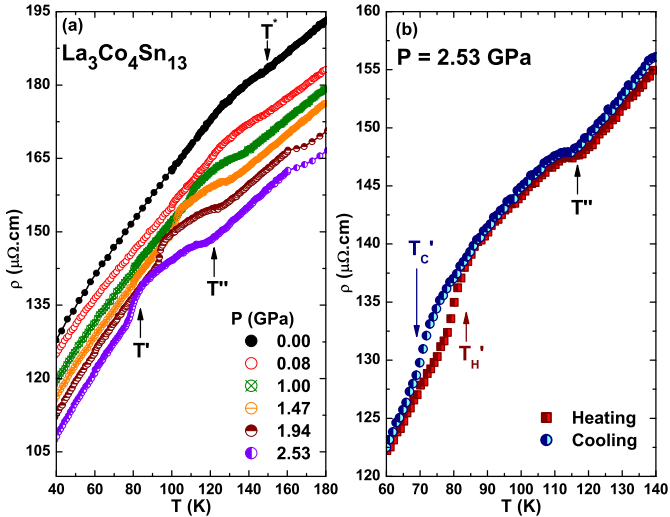}
\end{center}
\caption{(a) Representative pressure-dependent electrical resistivity curves as a function of temperature of the superlattice structural transition $T^*$ in La$_{3}$Co$_4$Sn$_{13}$. (b) Temperature region around $T^*$ on heating (square) and cooling (circle) at $P=2.53$ GPa.}
\label{FigRhoTstarP}
\end{figure}

While at low temperatures $T_{c}$ is enhanced by pressure, as seen above, at higher temperatures we observe a different behavior. Figure \ref{FigRhoTstarP}(a) displays the pressure evolution of the $\rho(T)$ data for La$_{3}$Co$_{4}$Sn$_{13}$ in the temperature range $40<T<180$ K. These curves are representative of the qualitative behavior for all applied pressures. We observe the suppression of the superlattice structural transition, like in Sr$_{3-x}$Ca$_x$Ir$_4$Sn$_{13}$ \cite{qcp3413,LaP}. In addition, a clear hump-like feature develops for $P\gtrsim0.6$ GPa, where two characteristic temperatures are identified, $T'$ and $T''$, as depicted in Fig. \ref{FigRhoTstarP}. Once the hump-like feature emerges, an evident thermal hysteresis [$\Delta T=4(2)$ K] is observed at $T'$, as exemplified in Fig. \ref{FigRhoTstarP}(b) for $P=2.53$ GPa. We noted that this thermal hysteresis is independent of the sweeping rate on cooling/heating the sample (not shown). This behavior was reproduced for more than one sample. The low temperature $T'$ on cooling ($T_C'$) and heating ($T_H'$) as a function of pressure are presented in Fig. \ref{Fig5}(b). Both $T_C'$ [$dT_C'/dP=-21(1)$ K/GPa] and $T_H'$ [$dT_H'/dP=-19(2)$ K/GPa] are reduced under pressure with very similar rate, though twice faster than $T''$ [$dT''/dP=-11(1)$ K/GPa] [see Fig. \ref{Fig5}(b)]. These characteristics temperatures were obtained from the deviation of linearity of $\rho$ in the $ln(1/\rho)$ versus $lnT$ plots as in Refs. \cite{lacosnX,LaP}. The pressure dependence of $T''$ is consistent with earlier reports of the superlattice structural transition $T^*$ in polycrystalline La$_{3}$Co$_{4}$Sn$_{13}$ \cite{LaP}, though no indication of the hump-like feature was reported in that sample under pressure. We performed $\rho(T)$ measurements on cooling and heating under an applied magnetic field $B=9$ T at $P=2.01$ GPa (not shown). No appreciable changes were observed, indicating that $T'$ and $T''$ do not have a magnetic origin nor are influenced by an external magnetic field. The physical origin of $T'$ and $T''$ is discussed in the section below.

\begin{figure}
\begin{center}
\includegraphics[width=0.5\textwidth]{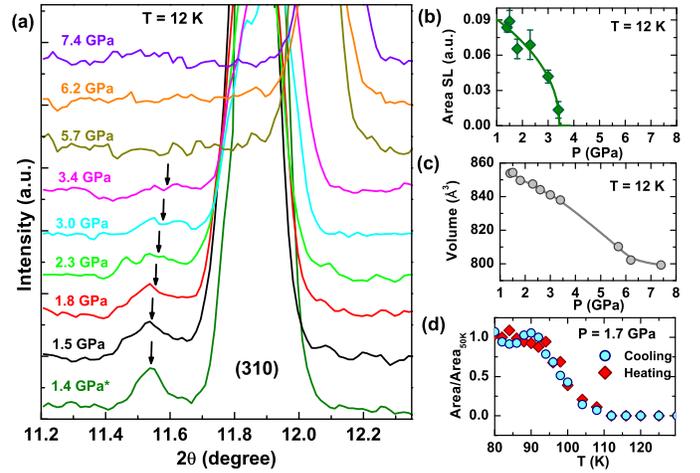}
\end{center}
\caption{(a) Pressure evolution of the superstructure reflection (3 0.5 0.5) at $T=12$ K in La$_{3}$Co$_4$Sn$_{13}$. $P=1.4$ GPa is after the DAC pressure was released. (b) Integrated area of the superlattice peak (3 0.5 0.5), at $T=12$ K, as a function of pressure. (c) Unit cell volume pressure dependence up to 7.4 GPa at $T=12$ K. (d) Temperature dependence of the superstructure reflection (3 0.5 0.5) integrated area at $P=1.7$ GPa, normalized at $T=50$ K. Solid lines are guide to the eyes.}
\label{FigSXRD2}
\end{figure}

It is not clear why $T^*$ develops into these two $T'$ and $T''$ characteristic temperatures. In order to elucidate the nature of these anomalies we performed synchrotron XRD under pressure. Figure \ref{FigSXRD2}(a) shows the pressure evolution of the superstructure reflections (2.5 1.5 1) or (3 0.5 0.5) at $T=12$ K, which has a well-defined peak at low pressures at around 2$\theta=11.54^\circ$. At this temperature, as pressure increases the peak broadens up to 3.4 GPa, and then disappears at higher pressures [Fig. \ref{FigSXRD2}(b)].  The vanishing of the superstructure reflection is accompanied by a huge contraction of the unit cell volume, as seen in Fig. \ref{FigSXRD2}(c), obtained by the (3 1 0) Bragg peak position data. Interestingly, after releasing the DAC pressure a residual $P=1.4$ GPa remained (checked by the ruby fluorescence line) and the superstructure peak reemerged at the same 2$\theta$ position with comparable width, indicating the process is reversible. Figure \ref{FigSXRD2}(d) displays the temperature dependence of the superstructure reflection (3 0.5 0.5) integrated area at $P=1.7$ GPa, normalized at $T=50$ K. We estimate, from the XRD data, the emergence of the superlattice structural transition at around $T=105(5)$ K ($P=1.7$ GPa), which is near $T'$ [see the pressure-temperature phase diagram, Fig. \ref{Fig5}(b)], although no thermal hysteresis is observed, in contrast to the $\rho(T)$ data.

\section{Discussion}

Our study shows that the pressure effect on single-crystalline La$_{3}$Co$_4$Sn$_{13}$, like in other 3-4-13 series of compounds, such as Sr$_{3-x}$Ca$_x$(Rh,Ir)$_4$Sn$_{13}$ \cite{qcp3413,qcp34132}, enhances the superconducting critical temperature and suppresses the superlattice structural transition. An indication of a quantum critical point, such as a linear temperature dependence of the electrical resistivity at low temperatures, however, is not observed up to $P=2.53$ GPa. Particularly for La$_{3}$Co$_4$Sn$_{13}$ under pressure, a hump-like feature emerges around the ambient pressure $T^*$, with two characteristic temperatures, observed in our $\rho(T)$ measurements. The low temperature $T'$ vanishes at a rate of $dT'/dP\approx-20$ K/GPa and we ascribe it as the onset of the superlattice structural transition, corroborated by our synchrotron XRD data. In addition, from $dT'/dP$ we estimate that the superlattice structural transition vanishes at $P\sim5.5$ GPa, and indeed no superstructure reflection peaks are seen for $P\geq5.7$ GPa at $T=12$ K [Fig. \ref{FigSXRD2}(a)].

The thermal hysteresis in the $\rho(T)$ data, around the superlattice structural transition, is not seen by cooling and heating the superstructure reflection [Fig. \ref{FigSXRD2}(d)]. Therefore, it is unlikely that the superlattice structural transition goes from a second (at ambient pressure) to a first order transition, for $P\gtrsim0.6$ GPa.  This thermal hysteresis, however, gives support to a CDW phase that sets in at $T^*$, indicating that the phenomenon seen in the electrical resistivity might be due to pinning of the CDW phase to impurities/defects, as observed in Lu$_{5}$Rh$_{4}$Si$_{10}$ \cite{pinCDW}.  The origin of the feature observed at the characteristic temperature $T''$, under pressure, might be related to the beginning of the CDW gap opening at higher temperatures and future studies should clarify this. Thus, a quantum critical point, which occurs only for vanishing second-order phase transitions, might still be observed at $P=5.5$ GPa \cite{mucio,RevLohneysen}.

In Sr$_{3-x}$Ca$_x$(Rh,Ir)$_4$Sn$_{13}$ compounds, at the superlattice structural transition, as the partially gapped CDW sets in, it results in an upturn on the $\rho(T)$ data \cite{qcp3413,qcp34132}. In La$_{3}$Co$_4$Sn$_{13}$, at ambient pressure, this upturn is less pronounced, pointing out that the size of the gap may be smaller. In addition, for our sample, the $dT_{c}/dP$ slope is one order of magnitude lower than the one reported for Ca$_3$Ir$_4$Sn$_{13}$ \cite{arxivca}, also allowing us to infer a smaller CDW gap. This is due to fact that pressure might promote the closing of the partially gapped CDW and the recovered $N(E_f)$ for La$_{3}$Co$_4$Sn$_{13}$ is possibly smaller than that for Ca$_3$Ir$_4$Sn$_{13}$, when the CDW is suppressed. Besides the rise in $T_{c}$, the CDW gap closure at the Fermi energy increases the carrier density, decreasing $\rho_0$ [Fig. \ref{FigRhoTcP}(a)], also evidencing the high quality of our single-crystalline samples. The interplay between the electronic effect and the quantum critical behavior to the enhancement of the superconducting critical temperature in 3-4-13 materials is not well established. For LuPt$_{2-x}$Pd$_x$In it has been observed that $T_{c}$ presents a dome-shaped doping dependence and its highest value is exactly where the CDW transition in this system disappears, and it was argued that in this case quantum fluctuations could play the major role \cite{CDWQCP}. Further experiments, such as $\rho(T)$ and specific heat, at higher pressures ($P\geq5.5$ GPa) are needed to clarify if La$_{3}$Co$_4$Sn$_{13}$ exhibit a quantum critical point and if it is relevant for superconductivity in this and other related materials.

\section{Summary}

In summary we performed $\rho(T)$ and XRD experiments on the superconducting La$_{3}$Co$_4$Sn$_{13}$ single crystals under pressure. We observe an enhancement of the superconducting transition temperature $T_c$ and the decrease of the superlattice structural transition $T^*$ as a function of pressure. This resembles the observed behavior of Sr$_{3-x}$Ca$_x$(Rh,Ir)$_4$Sn$_{13}$ similar compounds, in which a quantum critical point is reported where $T^*$ vanishes. For the La$_{3}$Co$_4$Sn$_{13}$ material we show that superlattice structural transition remains a second-order transition under pressure and estimate its suppression at $P\sim5.5$ GPa, though no quantum critical behavior is observed up to 2.53 GPa. Thermal hysteresis loops around $T^*$ are seen in the electrical resistivity curves for $P\gtrsim0.6$ GPa, and we argue that this effect is due to the pinning of a partially gapped CDW phase, which sets in at $T^*$. Altogether, our results demonstrate that the combination of transport and structural investigations under hydrostatic pressure can be extremely useful to understand the exotic physical proprieties of compounds such as the 3-4-13 family, and should guide the efforts to reveal the relation between structural and electronic properties in similar compounds with lattice instabilities.

\section*{Acknowledgments}
This work was supported by the Brazilian funding agencies: Funda\c{c}\~{a}o Carlos Chagas Filho de Amparo \`{a} Pesquisa do Estado do Rio de Janeiro (FAPERJ) [No. E-26/010.001045/2015], Conselho Nacional de Desenvlovimento Cient\'{\i}fico e Tecnol\'{o}gico (CNPq) [No. 400633/2016-7], and Funda\c{c}\~{a}o de Amparo \`{a} Pesquisa do Estado de S\~{a}o Paulo (FAPESP) [Nos. 2011/19924-2 and 2012/04870-7]. The authors thank the XDS-LNLS staff for technical support and LNLS for the for concession of beam time [proposal No. 20170611]. Part of this work was carried out at the light source PETRA III at DESY, a member of the Helmholtz Association (HGF). We would like to thank M. Ramakrishnan and M. A. Continentino for fruitful discussions and valuable suggestions.

\end{document}